\documentstyle[11pt,newpasp,twoside,epsf]{article}
\def\edcomment#1{\iffalse\marginpar{\raggedright\sl#1\/}\else\relax\fi}
\reversemarginpar

\begin{document}
\title{Stellar Populations, Butcher-Oemler Effect, Star Formation in Clusters}
 \author{Bianca M. Poggianti}
\affil{Osservatorio Astronomico di Padova-INAF, Italy}

\begin{abstract}
In this brief overview of the stellar populations of galaxies in
clusters I highlight some of the most recent results, including the
downsizing of the k+a population and the star formation--local density
relation.  I conclude discussing some open questions regarding the
interpretation of the observational results, and speculating on the
current meaning of the words ``primordial'' and ``environmental''.
\end{abstract}

\section{Introduction}

Studying the evolutionary histories of galaxies in clusters has always
been an endeavour to understand how galaxies (all galaxies) came to be
as we observe them. In recent years, this quest can be better
described as an effort to comprehend how galaxies formed and evolved
in a ``changing'' environment. In fact, the common framework for these
studies is now a hierarchical universe, in which galaxies can and
often change environment during their evolution. When a galaxy infalls
into a cluster, it can do so as a single isolated galaxy, as a pair,
as part of a group or as member of another merging cluster.  A galaxy
can experience very different environmental conditions throughout
its evolution. A major challenge for today's research is to identify
the effects produced by the various environments on the observable
properties of galaxies. It is especially hard to discriminate between
minor and major influences. If all environments are potentially
expected to have {\it some} impact on their galaxies, we would like to
discern secondary effects (``cosmetics'')
from the primary causes that determine how
galaxies are.

In this search we can rely on some objective and
measurable properties, such as mass in stars and gas, their chemical
content, their motion within the galaxy and the structures they
design.  The star formation activity is one of the most outstanding
characteristics.  Though a precise measurement of the current and past
star formation activity is no easy task for any galaxy, star formation
affects the accessible observables, such as colors and spectra, in such a
manifest way that is often the most immediate evidence for evolution. 

In this contribution I focus on the build-up of the
stellar populations in cluster galaxies and on the evolution with redshift
of their star formation activity.  The subject has been very rich in results
and papers published during the past 25 years, and it is impossible to
give an exhaustive summary in a short text. 
I will therefore only touch upon some of the issues, and highlight 
just a couple of the most recent results.
Overviews on the evolution of the stellar populations in cluster
galaxies 
can be found in Ellingson 2003, Poggianti 2003a and several reviews 
in the third volume of the Carnegie Observatories Astrophysics Series: Clusters of Galaxies: Probes of Cosmological Structure and Galaxy Evolution (eds. 
J.S. Mulchaey, A. Dressler, and A. Oemler (Pasadena: Carnegie Observatories,
http://www.ociw.edu/ociw/symposia/series/symposium3/proceedings.html). 
\section{Where it all started....and where we are}
The Butcher-Oemler effect is the excess of galaxies bluer than the
color-magnitude red sequence in clusters at $z>0.1-0.2$ as compared to
the richest nearby clusters (Butcher \& Oemler 1984).

The origin of the scatter in the fraction of blue galaxies from cluster
to cluster at any given redshift, and the consequent search for
correlations between the blue fraction $f_B$
and the global cluster properties (Smail et
al. 1998, Margoniner et al. 2001, Metevier et al. 2000);
the trend with redshift of $f_B$ in optical and X-ray selected
cluster samples and the dependence on cluster selection 
(Andreon \& Ettori 1999, Ellingson et
al. 2001, Kodama \& Bower 2001, Fairley et al. 2002); the 
change in galaxy colors with clustercentric distance (Pimbblet et
al. 2002); the nature of
the galaxies giving rise to the Butcher-Oemler effect (De Propris et
al. 2003); these are some of the aspects related to the Butcher-Oemler
effect that are still subject of ongoing investigation, witnessing
the interest in fully understanding the consequences and the cause
of the seminal Butcher \& Oemler findings.

Put in a slightly different way, 
the Butcher-Oemler effect is the presence of large numbers
of blue galaxies in rich clusters more distant than those in the local 
universe. This higher average level of activity in the past 
has been confirmed and greatly clarified by spectroscopic 
surveys and by Hubble Space Telescope morphological studies
of galaxies in distant clusters
(Couch et al. 1994, 1998, Dressler et al. 1997, Balogh et al. 1997, 
1998, 1999, Poggianti et al. 1999, van Dokkum et al. 1999, 2000, 
Fabricant et al. 2000, Ellingson et al. 2001, Postman et al. 2001, 
Smail et al. 2001, Lubin et al. 2002).

No matter how the evolution in the populations of cluster galaxies is
observed, if using the blue colors, the spectral features or the
proportion of spiral versus early-type galaxies: all the three types
of observations contribute to delineate a picture in which the
evolution of galaxies in clusters is strong, given that a large
fraction of them at z=0 have evolved from star-forming, late-type
galaxies to passively evolving, early-type galaxies within a
relatively short period of time (Kodama \& Smail 2001, van Dokkum 
\& Franx 2001).

At least three phenomena likely play a role in this evolution (Poggianti
et al. 1999, Ellingson et al. 2001, Kodama \& Bower 2001): the
declining infall rate of galaxies onto clusters at lower z predicted by 
hierarchical cosmological models (Bower 1991, Kauffmann 1995); 
the evolution with z of the average star formation rate (SFR) in the
``field'' galaxies that infall into clusters;
and the decline of star formation in the cluster galaxies likely
due to some physical process (or processes)
that acts when galaxies infall in the denser
environment. In fact, though the universe as a
whole seems to evolve towards a progressively lower star formation
activity in the Madau plot, such a trend appears to be accelerated in clusters
(Kodama \& Bower 2001).

\subsubsection{The past star formation activity as seen from the absorption lines}
The most cited result obtained from spectra of galaxies in distant clusters
is the presence of ``k+a'' spectra, with no emission lines and strong
Balmer lines in absorption indicating a vigorous star formation activity
that was terminated at some point during the last 1-1.5 Gyr
(Dressler \& Gunn 1983, Couch \& Sharples 1987).
Since such spectra inequivocally identify post-starburst or post-starforming
galaxies, their excess in distant clusters as compared to the field at similar
redshifts is strong, if not
the strongest, evidence that star formation is truncated by the dense
environment (Dressler et al. 1999, but see Balogh et al. 1999 for a 
different view). K+a galaxies have been thoroughly discussed elsewhere
(e.g. Poggianti 2003b and references therein), but it is worth 
highlighting here a very recent result from Tran et al. (2003).
On the basis of the galaxy velocity dispersions and radii, these authors find
that the descendants of the k+a's observed at $z\sim 0.8$ must be among
the most massive early-type galaxies in clusters today, while 
the maximum luminosity and galaxy velocity dispersion
of cluster k+a galaxies decrease towards lower redshifts. 
Another result pointing in the same direction is presented in 
Poggianti et al. (2003), where we analyzed the luminosity distribution of k+a
galaxies in the Coma cluster: no k+a as luminous as in the distant 
clusters is observed, while k+a's are found to be a significant fraction of
the dwarf galaxy population.
\subsubsection{The ongoing star formation activity as seen from the emission lines}

In the absence of a significant AGN ionizing radiation,
emission lines are present in a spectrum only when there is ongoing
star formation activity. 
\begin{figure}[h]
\plottwo{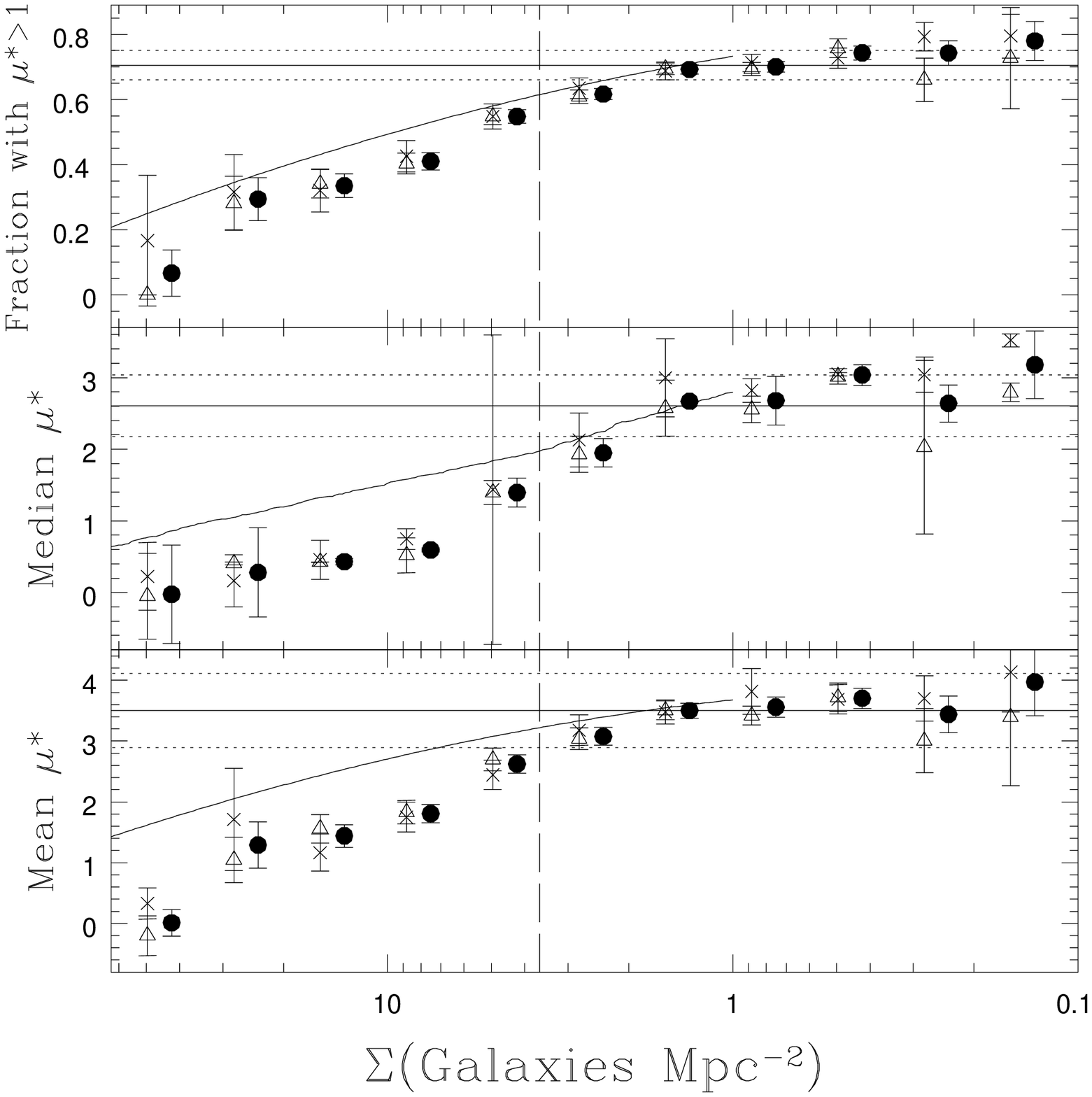}{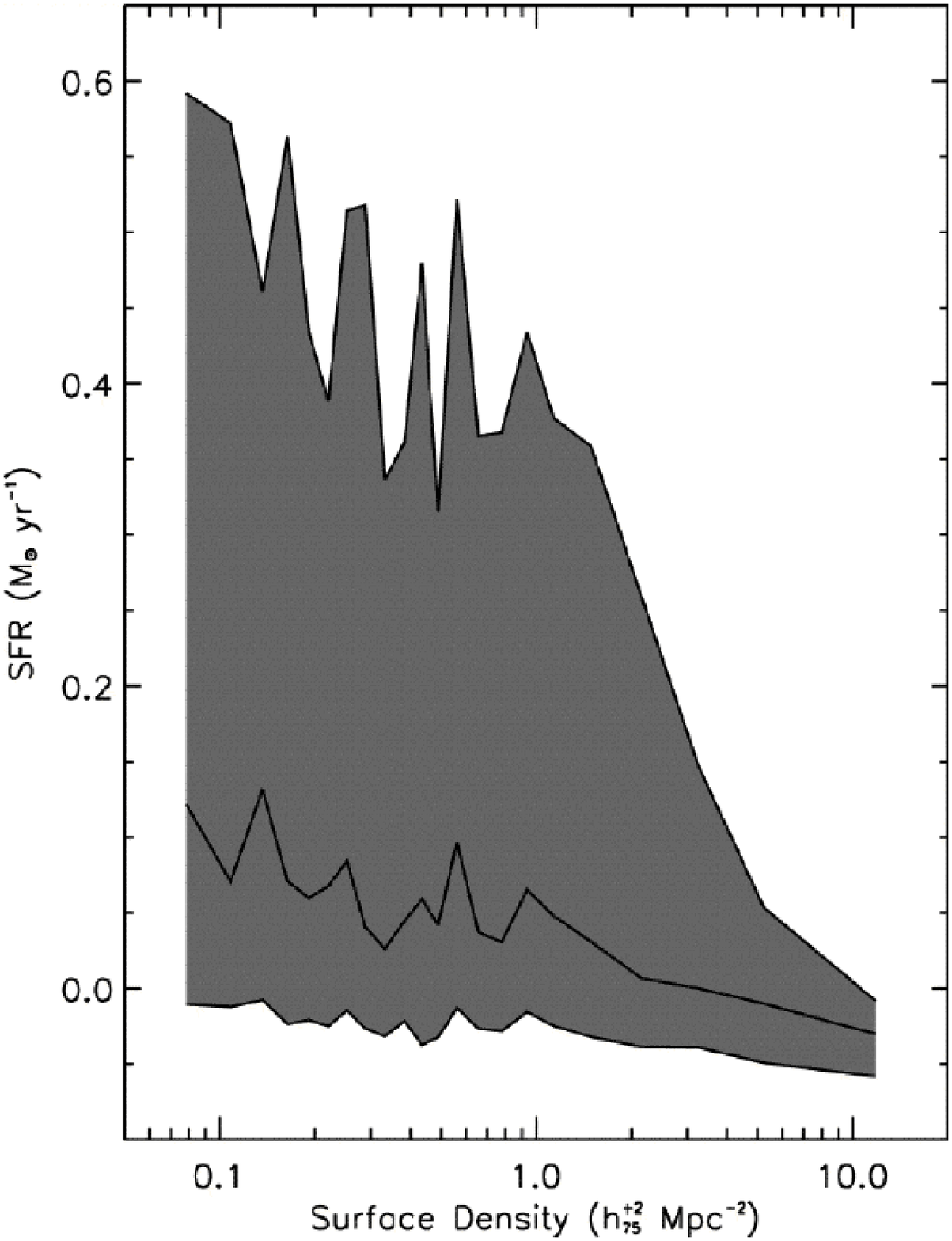}
\caption{{\bf Left} - From Lewis et al. 2002 (2dF): $\mu^*$, the star formation rate normalized to $L^*$, is proportional to the EW($\rm H\alpha$). Its trend as a function of the local galaxy density is shown for all clusters (solid points) and for clusters with $\sigma$ greater than or less than 800 $\rm km \, s^{-1}$ (triangles and crosses, respectively). The horizontal solid line represents the field value. {\bf Right} - From Gomez et al. 2003 (Sloan): the shaded 
area
represents the distribution of star formation rates, corrected for extinction. The median is shown as a solid line. The top of the shaded area is the $75^{th}$ percentile, while the bottom is the $25^{th}$ percentile.}
\end{figure}
In agreement with the other photometric and
spectroscopic signs of current activity, 
a significant fraction of even the most luminous galaxies in distant
clusters have been observed to have 
emission lines indicating they are actively forming
stars (Dressler et al. 1999, Postman et al. 2001), though the fraction of
emission line galaxies remains lower in clusters than in the field at
all redshifts. 

Whether the star formation activity in clusters is enhanced
in some of the infalling galaxies, and how relevant this is
for the global cluster populations is still debated.
Besides considerations based on optical observations
in favor and against cluster-triggered starbursts, a
striking result comes from dust-free wavelengths: the large numbers
 of Luminous Infrared starburst galaxies detected by ISOCAM in distant
clusters seem to be too high to be consistent with simple
accretion of starburst galaxies from the field
(Duc et al. submitted, Coia et al. submitted).

Finally, several works have been recently devoted to the study of the
emission line trend with clustercentric radius.  Balogh et al. (1997, 1998)
have shown how the mean equivalent width of [O~{\sc ii}] increases
radially in distant clusters, failing to reach the mean field value
even as far out as several Mpc from the cluster centre.  An extension
of these radial studies has been carried out using the large datasets
of the 2dF (Lewis et al. 2002) and Sloan (Gomez et al. 2003)
spectroscopic surveys. As shown in Fig.~1, the mean EW($\rm H\alpha$)
increases at lower projected local surface density of galaxies
(and, correspondingly, larger distance from cluster centre) down to
a density of $\sim 1$ galaxy per $\rm Mpc^2$, and approaches the field value
at  $\sim 3$ times the virial radius of the cluster.

The mean EW, or SFR, used in these studies is the mean over all
galaxies (with and without emission lines, of any morphological type),
including the passive early-type galaxies that dominate the cluster
cores and the highest density regions.  A mean total EW decreasing with
density and towards the cluster centre is therefore not surprising in
this respect. Whether the trend of mean EW {\it simply} reflects a
different morphological mix as a function of local density, and
whether it can be {\it fully} explained by the morphology-density
relation (Dressler et al. 1997) is still unclear (Lewis et al. 2002,
Gomez et al. 2003).

The {\it origin} of the observed EW trend remains in my opinion one of the
most interesting questions to be answered, together with the origin of
the morphology-density relation and the relation between the two.  A
systematic variation of galaxy properties with the environment does
not necessarily imply a ``transformation'' due to an environmental
process beginning to act on a galaxy when this enters an environment
for the first time. 
In principle, one cannot exclude that the correlation of star
formation with local density can be partly or fully explained as an
``imprinting'' on the galaxy star formation history  established in
the early universe as a function of the local density.  For example,
the most massive ellipticals in the densest regions today (the cluster
cores), that formed most of their stars at very high redshifts, were
in the densest regions also at $z>3$ and it is then, by those
conditions, that their star formation history was decided.  These
galaxies contribute to the shape of the EW-density relation today as
some of the galaxies with EW=0 (devoid of SF), and it is therefore
logical to conclude that at least {\it part} of the SF-density
relation has been established very early on.
Similarly, the morphology-density (MD) relation must have a
``primordial'' component, since at all redshifts the most massive
early-type galaxies are found in the highest density regions.
However, the MD relation as we observe it today is not {\it fully}
established at high redshift, because we observe how it evolves in
clusters, with late type spirals being transformed into early-type
galaxies (Dressler et al. 1997, Lubin et al. 2002).

It is probably time for the astronomical community working on clusters
to reconsider the meaning of words such as ``primordial'' and
``environmental''. By ``primordial'', do we mean anything that took
place at $z>3$, or anything that was already engraved in the galaxy
``destiny'' by the early conditions in the location where they formed
their first stars? And by ``environmental'', do we mean any physical
process that affects galaxies once they enter a ``new'' environment,
implying that their evolution would have been different if they did
not become part of it? Or do we consider an effect environmental any
time it has to do with something external to the galaxy itself? In
this case, what do we consider as the ``galaxy itself'', anything that
will be part of that galaxy by z=0?  Can a process be considered an
environmental effect even if it only affected the galaxy at $z>3$?  
Re-thinking the definition of the words ``primordial'' and 
``environmental'' is not
just a semantic exercise, it is probably useful to make progress in an
era in which we believe galaxies can change environment, but today's
environment is not necessarily unrelated with the environments of
yesterday.

\acknowledgments{I thank the IAU and the organizers of this Joint
Discussion on clusters for their kind invitation and for generously
supporting my participation with a IAU travel grant.}

\end{document}